\documentclass[a4paper,aps,prd,10pt,preprintnumbers,showkeys,showpacs,twocolumn,superscriptaddress,nofootinbib]{revtex4-1}\def\PRDonly#1{#1}\def\CQGonly#1{}
\usepackage{amssymb,orcidlink,graphicx,cmap}
\usepackage[utf8]{inputenc}
\usepackage[T1]{fontenc}

\CQGonly{%
\usepackage{cite,hyperref}
\bibliographystyle{unsrt}
\def\acknowledgments{\section*{Acknowledgements}}
}

\def\imo{i}
\def\Order#1{\mathcal{O}\left(#1\right)}
\def\re#1{\mathrm{Re}(#1)}
\def\im#1{\mathrm{Im}(#1)}
\newcommand{\eqref}[1]{(\ref{#1})}

\begin{document}

\title{WKB approximation for quasi-bound states and trapped modes}

\PRDonly{%
\author{R. A. Konoplya \orcidlink{0000-0003-1343-9584}}
\email{roman.konoplya@gmail.com}
\affiliation{Research Centre for Theoretical Physics and Astrophysics, Institute of Physics, Silesian University in Opava,\\ Bezručovo náměstí 13, CZ-74601 Opava, Czech Republic}
\author{Z. Stuchlík \orcidlink{0000-0003-2178-3588}}
\email{zdenek.stuchlik@physics.slu.cz}
\affiliation{Research Centre for Theoretical Physics and Astrophysics, Institute of Physics, Silesian University in Opava,\\ Bezručovo náměstí 13, CZ-74601 Opava, Czech Republic}
\author{A. Zhidenko \orcidlink{0000-0001-6838-3309}}
\email{olexandr.zhydenko@ufabc.edu.br}
\affiliation{Research Centre for Theoretical Physics and Astrophysics, Institute of Physics, Silesian University in Opava,\\ Bezručovo náměstí 13, CZ-74601 Opava, Czech Republic}
\affiliation{Centro de Matemática, Computação e Cognição (CMCC), Universidade Federal do ABC (UFABC),\\ Rua Abolição, CEP: 09210-180, Santo André, SP, Brazil}
}%

\CQGonly{%
\author{R. A. Konoplya \orcidlink{0000-0003-1343-9584},$^\dag$ Z. Stuchlík \orcidlink{0000-0003-2178-3588},$^\dag$ and A. Zhidenko \orcidlink{0000-0001-6838-3309}$^{\dag\ddag}$}
\address{$^\dag$ Research Centre for Theoretical Physics and Astrophysics, Institute of Physics, Silesian University in Opava, Bezručovo náměstí 13, CZ-74601 Opava, Czech Republic}
\address{$^\ddag$ Centro de Matemática, Computação e Cognição (CMCC), Universidade Federal do ABC (UFABC), Rua Abolição, CEP: 09210-180, Santo André, SP, Brazil}
\ead{olexandr.zhydenko@ufabc.edu.br}
}

\begin{abstract}
We develop a largely automatic semi-analytic method for calculating weakly damped quasi-bound states and trapped modes supported by a local minimum of an effective potential. The real part of the frequency is obtained from an arbitrarily high-order local WKB expansion and Padé resummation, while the exponentially small imaginary part is estimated using the Gamow approximation for tunnelling through one or two potential barriers. Because the local quantization requires only derivatives of the potential at its minimum, the method applies readily to non-rational effective potentials and to different compact-object geometries. We provide a Mathematica® notebook implementing the procedure. Comparisons with continued-fraction and direct numerical results for massive scalar fields around Schwarzschild and Kerr black holes and for axial trapped modes of a uniform-density star show that the method yields accurate real frequencies and useful estimates of decay rates outside the superradiant regime.
\end{abstract}

\PRDonly{
\maketitle
}%

\section{Introduction}

Massive fields propagating in black-hole spacetimes can form quasi-bound states, namely long-lived, spatially localized configurations \cite{Detweiler:1980uk,Dolan:2007mj} whose frequencies have real parts close to the mass threshold. Unlike quasinormal modes, which satisfy purely outgoing boundary conditions at spatial infinity and describe the prompt relaxation of perturbations, quasi-bound states have exponentially decaying spatial profiles. In the absence of superradiance, they gradually lose energy through absorption at the event horizon, giving a small but finite decay rate. Their existence results from the interplay between the gravitational attraction of the black hole and the confining effect of the field mass, while additional interactions \cite{Furuhashi:2004jk,Rosa:2011my}, such as electromagnetic coupling for charged fields, can substantially modify the spectrum, spatial localization, and lifetime of these states.

Quasi-bound states have attracted considerable attention owing to their connections with black-hole superradiance \cite{Starobinsky:1973a,Starobinsky:1973b,Brito:2015oca,Arvanitaki:2010sy,Konoplya:2011qq}, long-lived bosonic clouds, potential gravitational-wave signatures, constraints on ultralight particles beyond the Standard Model, and the late-time evolution of perturbations \cite{Hui:2016ltb}. Therefore, they provide a valuable and complementary probe of black-hole spacetimes, revealing aspects of the near-horizon geometry and field dynamics that are not directly accessible through the study of quasinormal modes alone.

While the energy levels and lifetimes of quasi-bound states can be calculated accurately using numerical methods such as Leaver's method \cite{Leaver:1985ax}, which is widely employed for computing quasinormal modes, these techniques also have certain limitations. In particular, Leaver's method is most straightforward when the radial equation has rational coefficients. Moreover, its implementation relies on an analysis of the singular points of the radial equation and is therefore highly case-dependent, so the corresponding Frobenius expansion often has to be modified for each problem. Another practical complication is that solving the continued-fraction equation numerically typically requires a sufficiently accurate initial guess for the complex frequency. Such a guess can often be obtained by continuously following a known mode as the parameters of the system are varied. However, this strategy may fail for non-perturbative branches of solutions that are not continuously connected to any known mode in a limiting regime. Locating such modes may then require an extensive search over the complex-frequency plane, substantially increasing the computational cost and introducing the risk that some roots will be overlooked.

The semi-analytic WKB method, which is widely used for calculating quasinormal modes \cite{Schutz:1985km,Iyer:1986np,Konoplya:2003ii,Matyjasek:2017psv}, can, in contrast, be automated for a wide class of effective potentials, although it is not completely universal \cite{Konoplya:2019hlu}. Importantly, it can also be applied to non-rational potentials. Another advantage is that the WKB expansion has recently been developed to arbitrarily high orders \cite{Matyjasek:2019eeu}, allowing high accuracy in suitable regimes, and an internal convergence criterion based on Cesàro means has also been formulated \cite{Konoplya:2026rjh}. We therefore use the higher-order WKB method to determine the energy levels of quasi-bound states with high accuracy. To estimate their decay rates, we employ the Gamow tunnelling expansion \cite{Gamow:1928zz}, in which the imaginary part of the frequency is treated as an exponentially small correction to the real WKB eigenvalue.

A related semiclassical treatment has been developed for trapped axial quasinormal modes of ultracompact stars \cite{Kokkotas:1994an}. In the weak-damping limit, the three-turning-point Bohr--Sommerfeld approach determines the real frequencies by quantizing the well action and the decay rates from tunnelling through the outer potential barrier \cite{Volkel:2017ofl}. These relations were subsequently used to reconstruct the trapping well and barrier from the axial spectra of constant-density stars and gravastars \cite{Volkel:2017kfj} and to address inverse problems in analogue-gravity systems \cite{Albuquerque:2024cwl}. The analogous four-turning-point construction has also been applied to wormholes with symmetric double-barrier potentials \cite{Volkel:2018hwb}. The present method retains this separation of oscillation and decay, but obtains the real frequency from the local higher-order WKB formula at the potential minimum rather than from an integral quantization condition.

The remainder of the paper is organized as follows. In Sec.~\ref{sec:boundstate-problem}, we formulate the quasi-bound-state problem for a massive scalar field propagating in a general static, spherically symmetric black-hole spacetime and introduce the analogous trapped-mode problem for compact stars and gravastars, together with the corresponding boundary conditions. Section~\ref{sec:WKBformula} presents the local higher-order WKB quantization condition near the potential minimum and its Padé-resummed version. In Sec.~\ref{sec:Gamow}, we derive the decay rate using the Gamow tunnelling expansion, while Sec.~\ref{sec:two-barriers} generalizes the formalism to the case of trapping wells separated from both asymptotic regions by potential barriers. In Sec.~\ref{sec:continued-fraction}, we briefly review the continued-fraction method for massive scalar perturbations of the Kerr black hole, which serves as a benchmark for our approach. Section~\ref{sec:numerical-comparison} compares the WKB results with accurate numerical calculations. Section~\ref{sec:higher-order} discusses our attempts to extend the decay-rate calculation to higher WKB orders, and Sec.~\ref{sec:conclusions} summarizes the conclusions.

\section{Quasi-bound states and trapped modes}\label{sec:boundstate-problem}

Consider a massive scalar field on a static, spherically symmetric black-hole background,
\begin{equation}
 ds^2=-f(r)dt^2+\frac{B^2(r)}{f(r)}dr^2+r^2(d\theta^2+\sin^2\theta d\phi^2),
 \label{eq:metric}
\end{equation}
where the event horizon is located at a zero of the blackening factor \(f(r)\), while \(B(r)\) is assumed to be regular there. The field obeys the Klein--Gordon equation
\begin{equation}
 \left(\Box-\mu^2\right)\Phi=0.
 \label{eq:kg}
\end{equation}
After separation of variables,
\begin{eqnarray}
 \Phi(t,r,\theta,\phi)
 &=&e^{-\imo\omega t}Y_{\ell m}(\theta,\phi)\frac{R(r)}{r},
 \label{eq:separation}
\end{eqnarray}
the radial equation can be written in the Schrödinger form
\begin{equation}
 \frac{d^2R}{dx^2}+\left[\omega^2-V(r)\right]R=0 ,
 \label{eq:radial}
\end{equation}
where \(x\) is the tortoise coordinate,
\begin{equation}\label{eq:tortoise}
 dx=\frac{B(r)}{f(r)}dr,
\end{equation}
and the effective potential is given by
\begin{equation}
 V(r)
 =f(r)\left[
 \mu^2+\frac{\ell(\ell+1)}{r^2}
 +\frac{1}{rB(r)}\frac{d}{dr}\left(\frac{f(r)}{B(r)}\right)
 \right] .
 \label{eq:potential}
\end{equation}

The qualitative shape of the effective potential is essential. At the horizon \(x\to-\infty\), the potential vanishes for an asymptotically flat black hole, so the local solutions are plane waves. At infinity, \(V(r)\to\mu^2\). Therefore, a massive field can support quasi-bound states with
\begin{equation}
 \omega^2<\mu^2 ,
\end{equation}
because the wave is exponentially suppressed at large radius. When the potential has an inner maximum and an outer minimum, a trapping well is formed outside the horizon. A level in this well is not perfectly bound, since it can tunnel through the inner barrier and be absorbed by the black hole.

With the time dependence \(e^{-\imo\omega t}\), the horizon boundary condition is ingoing for both quasinormal modes and quasi-bound states:
\begin{equation}
 R\propto e^{-\imo\omega x},\quad x\to-\infty.
 \label{eq:horizonbc}
\end{equation}
The distinction lies at infinity (\(x\to\infty\)). Quasinormal modes satisfy an outgoing-wave condition,
\begin{equation}
 R\propto e^{+\imo k x},\qquad k=\sqrt{\omega^2-\mu^2},
 \label{eq:qnmbc}
\end{equation}
with the branch chosen by analytic continuation \cite{Konoplya:2006br}. Quasi-bound states satisfy an exponentially decaying boundary condition,
\begin{equation}
 R\propto e^{-\eta x},\qquad
 \eta=\sqrt{\mu^2-\omega^2},\quad \re{\eta}>0 .
 \label{eq:boundbc}
\end{equation}
Thus the same radial equation defines two different spectral problems. The absorbing horizon makes the quasi-bound-state frequencies complex, with \(\im{\omega}<0\) for stable Schwarzschild quasi-bound states.

Another closely related problem arises for horizonless ultracompact objects. For perturbations of sufficiently compact stars and gravastars, the effective potential may contain a local minimum separated from spatial infinity by an outer potential barrier \cite{Volkel:2017ofl,Volkel:2017kfj}. The modes localized in this well are trapped modes. Their real frequencies are determined mainly by oscillations near the minimum, whereas their decay is governed by tunnelling through the barrier and the subsequent emission of radiation to infinity. The mechanism is therefore analogous to that of a massive-field quasi-bound state: in both cases a potential well determines the approximately bound spectrum, while weak leakage through a barrier makes the frequencies complex with small imaginary parts. The principal difference is that a black-hole quasi-bound state tunnels through the inner barrier and is absorbed at the event horizon while being exponentially suppressed at infinity, whereas a stellar or gravastar trapped mode leaks outward through the outer barrier. Such a trapped mode is consequently a quasinormal mode of the horizonless object rather than a quasi-bound state.

With the time dependence \(e^{-\imo\omega t}\), the trapped-mode problem is defined by regularity at the centre and a purely outgoing condition at spatial infinity. Denoting the corresponding radial master function by the same symbol \(R\), these boundary conditions can be written as
\begin{eqnarray}
 R\propto r^{\ell+1},&\quad& r\to0,
 \label{eq:centerbc}
 \\
 R\propto e^{+\imo\omega x},&\quad& x\to+\infty.
\end{eqnarray}
For an ordinary compact star, the interior solution is matched to the exterior solution at the stellar surface. When the effective potential has no singular contribution there, both \(R\) and \(dR/dx\) are continuous. For a gravastar, regularity at the centre and outgoing behavior at infinity must be supplemented by the appropriate junction conditions at each material interface or thin shell. In both cases, the local WKB construction used below applies directly when the effective potential is smooth near its nondegenerate minimum.

\section{Local two-turning-point WKB formula}\label{sec:WKBformula}

The standard higher-order WKB formula is derived by expanding the radial equation near an extremum \(x_m\) of the effective potential \eqref{eq:potential}.
Following Ref.~\cite{Matyjasek:2026yiu}, we define
\begin{equation}
 V_k=
 \frac{d^kV}{dx^k}\Biggr|_{x=x_m}
 \qquad (V_1=0).
 \label{eq:Vderivatives}
\end{equation}
Near the extremum, the two closest turning points are treated uniformly by matching the WKB solutions to parabolic-cylinder functions. This removes the singular behavior of the naive WKB expansion at the turning points and gives an algebraic relation between \(\omega\), an auxiliary WKB quantity \(K\), and the derivatives \(V_k\).

In its generic form, the result can be written as
\begin{eqnarray}
 \omega^2
 &=&V_0+A_2(K^2)+A_4(K^2)+A_6(K^2)+\cdots
 \\\nonumber
 &&{}-\imo K\sqrt{-2V_2}
 \left[1+A_3(K^2)+A_5(K^2)+\cdots\right],
 \label{eq:genericwkb}
\end{eqnarray}
where \(A_k(K^2)\) denotes the WKB correction of order \(k\).
These corrections are polynomials in \(K^2\) and in the derivatives \(V_2,V_3,\ldots,V_{2k}\), divided by appropriate powers of \(V_2\).
For quasinormal modes one imposes the pole condition
\begin{equation}
 K=n+\frac12,\qquad n=0,1,2,\ldots ,
 \label{eq:Kqnm}
\end{equation}
for \(\re{\omega}>0\), and then solves Eq.~\eqref{eq:genericwkb} for the complex frequency.

\begin{figure}
\CQGonly{\centering}
\PRDonly{\resizebox{\linewidth}{!}}{\includegraphics{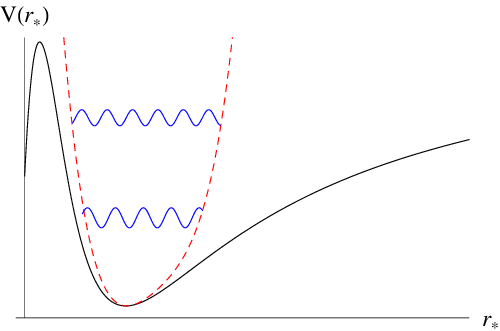}}
\caption{Schematic effective potential for a massive scalar field in the Schwarzschild background. The red dashed curve represents the local approximation to the trapping well, and the blue curves show the corresponding bound states.}
\label{fig:potential}
\end{figure}

When the formula is applied at a local maximum, \(V_2<0\), it yields the familiar two-turning-point approximation for barrier resonances and quasinormal modes (see Fig.~\ref{fig:potential}). For quasi-bound states, one instead applies the same local expression at the outer minimum, where \(V_2>0\). In this case the two turning points closest to the minimum bound the local trapping well. After the appropriate redefinition of the local coordinate, the outgoing-wave sectors of the barrier problem are mapped to the two decaying WKB tails of the bound-state problem. Therefore, the same algebraic WKB formula can be used, with \(V_2>0\), to obtain the local spectrum in the well. Equivalently, \(\sqrt{-2V_2}\) is continued through the complex plane, so the damping term for barrier resonances becomes the real oscillator spacing of the bound-state spectrum \cite{Zaslavsky:1991ug}.

Since the WKB series is generally asymptotic, increasing its order does not necessarily improve the result. Following Ref.~\cite{Konoplya:2019hlu}, the truncated series can be reorganized using Padé approximants. At WKB order \(k\), one first introduces a formal order-counting parameter \(\epsilon\) and defines the polynomial
\begin{eqnarray}
 P_k(\epsilon)&=&V_0+A_2(K^2)\epsilon^2
 +A_4(K^2)\epsilon^4+A_6(K^2)\epsilon^6+\cdots
 \nonumber\\
 &&{}-\imo K\sqrt{-2V_2}\bigl[\epsilon+A_3(K^2)\epsilon^3
 +A_5(K^2)\epsilon^5+\cdots\bigr],
 \label{eq:wkbpolynomial}
\end{eqnarray}
where all terms through order \(\epsilon^k\) are retained. Thus the ordinary truncated WKB result is \(\omega^2=P_k(1)\).

The Padé approximant associated with the same WKB order is the rational function
\begin{equation}
 P_{\widetilde n/\widetilde m}(\epsilon)
 =\frac{Q_0+Q_1\epsilon+\cdots+Q_{\widetilde n}\epsilon^{\widetilde n}}
 {R_0+R_1\epsilon+\cdots+R_{\widetilde m}\epsilon^{\widetilde m}},
 \quad \widetilde n+\widetilde m=k,
 \label{eq:wkbpade}
\end{equation}
whose coefficients are fixed by the matching condition
\begin{equation}
 P_{\widetilde n/\widetilde m}(\epsilon)-P_k(\epsilon)
 =\Order{\epsilon^{k+1}}.
 \label{eq:wkbpadematching}
\end{equation}
The Padé-resummed squared frequency is then
\begin{equation}
 \omega^2=P_{\widetilde n/\widetilde m}(1).
 \label{eq:wkbpadefrequency}
\end{equation}
For the local minimum, the analytic continuation of \(\sqrt{-2V_2}\) described above is understood in Eq.~\eqref{eq:wkbpolynomial}. The Padé resummation therefore reorganizes the same local WKB data and does not change the bound-state condition. Near-diagonal approximants, \(\widetilde n\simeq\widetilde m\), are usually the most accurate, although nearby choices should be compared because Padé resummation does not guarantee convergence. In particular, we use the approximation \(P_{8/8}(1)\), corresponding to \(k=16\).

This construction solves only the local bound-state quantization problem in the trapping well. Since the local approximation replaces the exterior by an effectively infinite well, the WKB formula does not impose the ingoing condition at the horizon. Consequently, it gives only an approximation to the real part of \(\omega\). To obtain the imaginary part, which corresponds to the decay rate of the mode, tunnelling through the potential barrier must also be taken into account.

\section{Decay rate from the Gamow tunnelling expansion}\label{sec:Gamow}

Let \(\omega_R^2\) be obtained from the local WKB formula at the minimum. If \(\omega_R^2\) lies below the barrier top and below the mass threshold, the equation
\begin{equation}
V(x)=\omega_R^2
\end{equation}
has three real turning points, \(x_1<x_2<x_3\). The region \(x_1<x<x_2\) is the forbidden barrier separating the well from the horizon, while \(x_2<x<x_3\) is the classically allowed well.

The leading Gamow approximation treats the imaginary part as a small nonperturbative correction to the real level \cite{Gamow:1928zz}. We define the barrier action and the well action as follows:
\begin{equation}
\begin{array}{rcl}
 K(\omega_R^2)&=&\displaystyle{\intop_{x_1}^{x_2}\sqrt{V(x)-\omega_R^2}\,dx},\\
 I(\omega_R^2)&=&\displaystyle{\intop_{x_2}^{x_3}\sqrt{\omega_R^2-V(x)}\,dx}.
\end{array}
 \label{eq:actions}
\end{equation}

The factor \(\exp[-2K]\) is the leading tunnelling probability through the inner barrier. The well-normalization factor is determined by
\begin{equation}
 I'(\omega_R^2)=\frac{1}{2}\intop_{x_2}^{~~x_3}
 \frac{dx}{\sqrt{\omega_R^2-V(x)}},
 \label{eq:Iderivative}
\end{equation}
This derivative sets the semiclassical timescale for repeated encounters of the particle with the well boundaries.

The three-turning-point connection formula gives the imaginary part of
\(\omega^2\) \cite{Patrick:2021oqk},
\begin{equation}
 \mathrm{Im}\,\omega^2
 = -\frac{1}{2I'(\omega_R^2)}
 \ln\frac{4+\exp[-2K(\omega_R^2)]}{4-\exp[-2K(\omega_R^2)]},
 \label{eq:imagE}
\end{equation}
which, to leading order in the tunnelling probability, yields
\begin{equation}
 \omega
 =
 \omega_R
 -\imo\,\frac{\exp[-2K(\omega_R^2)]}{8\omega_R I'(\omega_R^2)}
 +\Order{\omega_R^3}.
 \label{eq:imagomegaexpanded}
\end{equation}

In the radial coordinate, the same quantities are evaluated using \eqref{eq:tortoise}. Thus,
\begin{eqnarray}
 K(\omega_R^2)&=&\intop_{r_1}^{~~r_2}
 \frac{B(r)\sqrt{V(r)-\omega_R^2}}{f(r)}\,dr,\\
 I'(\omega_R^2)&=&\frac12\intop_{r_2}^{~~r_3}
 \frac{B(r)\,dr}{f(r)\sqrt{\omega_R^2-V(r)}} .
 \label{eq:radialactions}
\end{eqnarray}

The same construction provides an approximate solution of the stellar trapped-mode problem~\cite{Volkel:2017ofl,Volkel:2017kfj}. The axial perturbation equation has the form of Eq.~\eqref{eq:radial}, with the appropriate stellar effective potential. Regularity at the centre replaces the ingoing horizon condition, while the wave is outgoing at infinity. For \(x_1<x_2<x_3\), the allowed well now occupies \(x_1<x<x_2\), and the forbidden barrier occupies \(x_2<x<x_3\). Accordingly, \(I\) and \(I'\) in Eqs.~\eqref{eq:actions} and \eqref{eq:Iderivative} are evaluated between \(x_1\) and \(x_2\), while \(K\) is evaluated between \(x_2\) and \(x_3\).

One can thus determine \(\omega_R\) using the local WKB formula and its Padé resummation from Sec.~\ref{sec:WKBformula}, evaluated at the minimum of the stellar potential, and estimate the imaginary part from Eq.~\eqref{eq:imagomegaexpanded} with these integration limits. The resulting frequency describes a quasinormal mode of the star, with damping caused by radiation leaking through the outer barrier to infinity. The method applies to weakly damped modes localized near a smooth, nondegenerate minimum. A cusp or thin shell at the minimum requires separate matching and cannot be treated directly with the derivative-based WKB formula.

\section{Decay through two barriers}\label{sec:two-barriers}

A related situation occurs when the effective potential decreases toward the asymptotic regions rather than approaching the mass-squared limit, as happens, for example, in asymptotically de Sitter spacetimes. Then a potential well may be formed between two peaks. The corresponding mode is a quasinormal mode rather than a quasi-bound state: the trapped wave can leak through both barriers and escape toward both asymptotic regions. Nevertheless, when the leakage is weak, the real and imaginary parts of the frequency can be calculated in a closely analogous way.

Suppose that the equation
\begin{equation}
 V(x)=\omega_R^2
\end{equation}
has four real turning points, \(x_1<x_2<x_3<x_4 \). The classically allowed well is \(x_2<x<x_3\). The intervals \(x_1<x<x_2\) and \(x_3<x<x_4\) are the left and right forbidden barriers, respectively. The real part of the frequency is again determined by the WKB approximation at the local minimum.

The two barrier actions are
\begin{equation}
\begin{array}{rcl}
 K_L(\omega_R^2)&=&\displaystyle{\intop_{x_1}^{~~x_2}
 \sqrt{V(x)-\omega_R^2}\,dx},\\[1ex]
 K_R(\omega_R^2)&=&\displaystyle{\intop_{x_3}^{~~x_4}
 \sqrt{V(x)-\omega_R^2}\,dx}.
\end{array}
 \label{eq:doublebarrieractions}
\end{equation}
If both barriers are sufficiently opaque, the two leakage channels contribute
additively to the decay rate,
\begin{equation}
 \omega
 \simeq
 \omega_R
 -\imo\,\frac{\exp[-2K_{L}(\omega_R^2)]+\exp[-2K_{R}(\omega_R^2)]}{8\omega_R I'(\omega_R^2)}.
 \label{eq:doubleimagomegaexpanded}
\end{equation}
Thus each barrier produces a partial decay width, and the total width is the sum of the two partial widths at this order. If one barrier is more transparent than the other, it dominates the imaginary part.

This construction also applies to regular wormholes whose effective potential has two barriers enclosing a well around the throat. The potential \(V(x)\) must be smooth in the tortoise coordinate across the throat, with a nondegenerate minimum and finite derivatives up to the order required by the WKB approximation.

\section{Massive scalar field in Kerr and the continued-fraction method}\label{sec:continued-fraction}

To test the WKB approximation, we compare it with accurate frequencies obtained using the Frobenius continued-fraction method for a massive scalar field in the Kerr spacetime \cite{Dolan:2007mj}.

In Boyer--Lindquist coordinates, the Kerr line element is
\begin{eqnarray}
 ds^2&=&-\left(1-\frac{2Mr}{\Sigma}\right)dt^2-\frac{4Mar\sin^2\theta}{\Sigma}\,dt\,d\phi+\frac{\Sigma}{\Delta}\,dr^2
 \nonumber\\
 &&{}+\Sigma\,d\theta^2
 +\frac{\left[(r^2+a^2)^2-a^2\Delta\sin^2\theta\right]
 \sin^2\theta}{\Sigma}\,d\phi^2,
 \label{eq:kerrmetric}
\end{eqnarray}
where
\begin{equation}
 \Sigma=r^2+a^2\cos^2\theta,
 \qquad
 \Delta=r^2-2Mr+a^2,
 \label{eq:kerrmetricfunctions}
\end{equation}
and \(a=J/M\) is the specific angular momentum of the black hole.

The Klein--Gordon equation~\eqref{eq:kg} is separated using the ansatz
\begin{equation}
 \Phi=e^{-\imo\omega t+\imo m\phi}S_{\ell m}(\theta)R_{\ell m}(r).
 \label{eq:kerrseparation}
\end{equation}
The resulting angular and radial equations are
\begin{eqnarray}
 &&\frac{1}{\sin\theta}\frac{d}{d\theta}
 \left(\sin\theta\frac{dS_{\ell m}}{d\theta}\right)
 \PRDonly{
 \nonumber\\
 &&\quad
 }%
 +\left[a^2(\mu^2-\omega^2)\sin^2\theta
 -\frac{m^2}{\sin^2\theta}+\lambda_{\ell m}\right]S_{\ell m}=0,
 \label{eq:kerrangular}\\
 &&\frac{d}{dr}\left(\Delta\frac{dR_{\ell m}}{dr}\right)
 +\left[\frac{K^2(r)}{\Delta}-\mu^2r^2-\Lambda_{\ell m}\right]R_{\ell m}=0,
 \label{eq:kerrradial}\\
 &&K(r)=(r^2+a^2)\omega-am,
 \nonumber\\
 &&\Lambda_{\ell m}=\lambda_{\ell m}-2am\omega+a^2\mu^2.
 \nonumber
\end{eqnarray}
Here \(\lambda_{\ell m}\) is the scalar spheroidal eigenvalue and tends to \(\ell(\ell+1)\) as \(a\to0\). In Mathematica, it is obtained from
\begin{equation}
 \lambda_{\ell m}=\mathop{\mathrm{SpheroidalEigenvalue}}
 \left[\ell,m,\imo a\sqrt{\omega^2-\mu^2}\right].
 \label{eq:kerrangularmathematica}
\end{equation}

The radial solution satisfying the ingoing condition at the horizon and the decaying condition at infinity is written as
\begin{eqnarray}
 R_{\ell m}(r)&=&(r-r_+)^{-\imo\sigma}(r-r_-)^{\imo\sigma+\chi-1}e^{-\eta r}
 \PRDonly{
 \nonumber\\
 &&{}\times
 }%
 \sum_{n=0}^{\infty}a_n\left(\frac{r-r_+}{r-r_-}\right)^n,
 \label{eq:kerrfrobeniusseries}
\end{eqnarray}
where \(r_+\) and \(r_-\) are the event and inner horizons, respectively:
$$
 r_\pm=M\pm\sqrt{M^2-a^2}, \qquad r_++r_-=2M, \quad r_+r_-=a^2.
$$
We also define
\begin{equation}
 \eta=\sqrt{\mu^2-\omega^2},
 \quad \re{\eta}>0,
 \label{eq:kerrfrobeniusdefinitions}
\end{equation}
and
\begin{equation}
\begin{array}{rl}
 \sigma={}&\displaystyle\frac{K(r_+)}{r_+-r_-}
 =\displaystyle\frac{r_+(r_++r_-)\omega-am}{r_+-r_-},
\\
 \chi={}&\displaystyle\frac{(r_++r_-)(2\omega^2-\mu^2)}{2\eta}.
\end{array}
\label{eq:kerrfrobeniusexponents}
\end{equation}

Substitution of Eq.~\eqref{eq:kerrfrobeniusseries} into Eq.~\eqref{eq:kerrradial} gives the three-term recurrence
\begin{eqnarray}
 \alpha_0a_1+\beta_0a_0&=&0,\nonumber\\
 \alpha_na_{n+1}+\beta_na_n+\gamma_na_{n-1}&=&0,
 \qquad n\geq1,
 \label{eq:kerrrecurrence}
\end{eqnarray}
with
\begin{equation}
\begin{array}{rcl}
 \alpha_n&=&n^2+(c_0+1)n+c_0,\\
 \beta_n&=&-2n^2+(c_1+2)n+c_3,\\
 \gamma_n&=&n^2+(c_2-3)n+c_4,
\end{array}
 \label{eq:kerrrecurrencecoefficients}
\end{equation}
where, following Ref.~\cite{Dolan:2007mj}, we define
\begin{eqnarray}
 c_0&=&1-2\imo\sigma,\nonumber\\
 c_1&=&-4-4r_+\eta+\frac{(r_++r_-)\mu^2}{\eta}
 +4\imo\sigma,\nonumber\\
 c_2&=&3-2\chi-2\imo\sigma,\nonumber\\
 c_3&=&r_+(r_++r_-)(\omega+\imo\eta)^2
 -2am(\omega+\imo\eta)
 \nonumber\\
 &&{}+r_+r_-\mu^2-\Lambda_{\ell m}-1
 -(r_+-r_-)\eta+\chi
 \PRDonly{
 \nonumber\\
 &&{}
 }%
 +2\imo\left[1-\chi-\imo(r_++r_-)\omega\right]\sigma,
 \nonumber\\
 c_4&=&\left[\frac{(r_++r_-)\mu^2}{2\eta}\right]^2
 +2\left[\imo\chi-(r_++r_-)\omega\right]\sigma.
 \nonumber
\end{eqnarray}
The quasi-bound-state frequencies are the roots of the infinite continued-fraction equation
\begin{equation}
 0=\beta_0-
 \frac{\alpha_0\gamma_1}{\displaystyle
 \beta_1-\frac{\alpha_1\gamma_2}{\displaystyle
 \beta_2-\ldots}}.
 \label{eq:kerrcontinuedfraction}
\end{equation}
For each trial complex frequency, we first calculate \(\eta\) and the corresponding angular eigenvalue \(\lambda_{\ell m}\), which together determine all the recurrence coefficients. The continued fraction is then truncated at a sufficiently large index and evaluated backwards from the truncation point.

The WKB approximation also simplifies the calculation of accurate Frobenius frequencies by supplying an initial guess for the numerical solution of Eq.~\eqref{eq:kerrcontinuedfraction}. For a chosen overtone, the local WKB formula determines the approximate real part, while the Gamow expression~\eqref{eq:imagomegaexpanded} supplies an estimate of the imaginary part on the decaying branch. When this complex estimate is sufficiently close to the desired root, a local root-finding iteration can refine it directly, thereby avoiding a global search over the complex-frequency plane and reducing the risk that the iteration converges to a different mode. Convergence is checked by monitoring the residual and increasing the truncation order.

\section{Numerical comparison}\label{sec:numerical-comparison}

To apply the WKB approximation, we rewrite the radial equation \eqref{eq:kerrradial} in a wave-like form. The tortoise coordinate is defined as
\begin{equation}
 dx=\frac{r^2+a^2}{\Delta}dr,
\end{equation}
which corresponds to \(f(r)=\Delta/(r^2+a^2)\) and \(B(r)=1\) in the notation of Eq.~\eqref{eq:tortoise}. The rescaled radial function is
\begin{equation}
 \Psi_{\ell m}(r)=\sqrt{r^2+a^2}\,R_{\ell m}(r).
\end{equation}

Equation~\eqref{eq:kerrradial} then takes the form
\begin{equation}
 \frac{d^2\Psi_{\ell m}}{dx^2}+\left[\omega^2-V(r,\omega)\right]\Psi_{\ell m}=0,
 \label{eq:kerrschrodinger}
\end{equation}
where the effective potential is
\begin{eqnarray}\label{eq:kerrpotential}
 &&V(r,\omega)=\omega^2-\frac{K^2(r)}{(r^2+a^2)^2}
 \\\nonumber&&\PRDonly{\!\!\!\!\!\!\!\!\!\!}
 +\frac{\Delta}{(r^2+a^2)^2}
 \left(\Lambda_{\ell m}+\mu^2r^2+r\frac{d}{dr}\frac{\Delta}{r^2+a^2}+\frac{a^2\Delta}{(r^2+a^2)^2}\right).
\end{eqnarray}

\begin{table}
\CQGonly{\flushright}
\PRDonly{\resizebox{\linewidth}{!}}{%
\begin{tabular}{|c|c|c|}
\hline
\hline
$\mu$ & WKB & Frobenius \\
\hline
%0.01 & $0.00999987\!-\!-8.88049\times10^{-21} \imo$ & $0.00999987\!-\!8.39054\times10^{-22} \imo$ \\
0.02 & $0.0199990\!-\!4.10079\times10^{-18} \imo$ & $0.0199990\!-\!8.74469\times10^{-19} \imo$ \\
%0.03 & $0.0299966\!-1.57595\times10^{-16} \imo$ & $0.0299966\!-\!5.18377\times10^{-17} \imo$ \\
0.04 & $0.0399920\!-\!2.19261\times10^{-15} \imo$ & $0.0399920\!-\!9.54914\times10^{-16} \imo$ \\
%0.05 & $0.0499843\!-1.75035\times10^{-14} \imo$ & $0.0499843\!-9.30444\times10^{-15} \imo$ \\
0.06 & $0.0599729\!-\!9.84144\times10^{-14} \imo$ & $0.0599729\!-\!6.07592\times10^{-14} \imo$ \\
%0.07 & $0.0699569\!-4.34727\times10^{-13} \imo$ & $0.0699569\!-\!3.01634\times10^{-13} \imo$ \\
0.08 & $0.0799356\!-\!1.61026\times10^{-12} \imo$ & $0.0799356\!-\!1.22719\times10^{-12} \imo$ \\
%0.09 & $0.0899081\!-5.21589\times10^{-12} \imo$ & $0.0899081\!-\!4.29424\times10^{-12} \imo$ \\
0.10 & $0.0998736\!-\!1.52037\times10^{-11} \imo$ & $0.0998736\!-\!1.33563\times10^{-11} \imo$ \\
%0.11 & $0.109831  -4.07015\times10^{-11} \imo$ & $0.109831-3.77966\times10^{-11} \imo$ \\
0.12 & $0.119781  -\!1.01588\times10^{-10} \imo$ & $0.119781-\!9.90006\times10^{-11} \imo$ \\
%0.13 & $0.129720  -2.39110\times10^{-10} \imo$ & $0.129720-2.43148\times10^{-10} \imo$ \\
0.14 & $0.139649  -\!5.35443\times10^{-10} \imo$ & $0.139649-\!5.65588\times10^{-10} \imo$ \\
%0.15 & $0.149567  -1.14876\times10^{-9}  \imo$ & $0.149567-1.25588\times10^{-9}  \imo$ \\
0.16 & $0.159473  -2.37458\times10^{-9}  \imo$ & $0.159473-2.67889\times10^{-9}  \imo$ \\
%0.17 & $0.169366  -4.75101\times10^{-9}  \imo$ & $0.169366-5.51753\times10^{-9}  \imo$ \\
0.18 & $0.179244  -9.23593\times10^{-9}  \imo$ & $0.179244-1.10190\times10^{-8}  \imo$ \\
%0.19 & $0.189106  -1.75021\times10^{-8}  \imo$ & $0.189106-2.14122\times10^{-8}  \imo$ \\
0.20 & $0.198953  -3.23737\times10^{-8}  \imo$ & $0.198953-4.06044\times10^{-8}  \imo$ \\
%0.21 & $0.208781  -5.87555\times10^{-8}  \imo$ & $0.208781-7.53266\times10^{-8}  \imo$ \\
0.22 & $0.218591  -1.04603\times10^{-7}  \imo$ & $0.218591-1.36992\times10^{-7}  \imo$ \\
%0.23 & $0.228380  -1.83073\times10^{-7}  \imo$ & $0.228380-2.44676\times10^{-7}  \imo$ \\
0.24 & $0.238147  -3.15406\times10^{-7}  \imo$ & $0.238147-4.29824\times10^{-7}  \imo$ \\
%0.25 & $0.247892  -5.35399\times10^{-7}  \imo$ & $0.247891-7.43628\times10^{-7}  \imo$ \\
0.26 & $0.257613  -8.96617\times10^{-7}  \imo$ & $0.257610-1.26840\times10^{-6}  \imo$ \\
%0.27 & $0.267306  -1.48383\times10^{-6}  \imo$ & $0.267302-2.13490\times10^{-6}  \imo$ \\
0.28 & $0.277024  -2.38704\times10^{-6}  \imo$ & $0.276964-3.54835\times10^{-6}  \imo$ \\
%0.29 & $0.286616  -3.91831\times10^{-6}  \imo$ & $0.286595-5.82667\times10^{-6}  \imo$ \\
0.30 & $0.296219  -6.27991\times10^{-6}  \imo$ & $0.296192-9.45565\times10^{-6}  \imo$ \\
%0.31 & $0.305787  -9.96955\times10^{-6}  \imo$ & $0.305752-1.51660\times10^{-5}  \imo$ \\
0.32 & $0.315313  -1.56978\times10^{-5}  \imo$ & $0.315272-2.40372\times10^{-5}  \imo$ \\
%0.33 & $0.324787  -2.45476\times10^{-5}  \imo$ & $0.324748-3.76302\times10^{-5}  \imo$ \\
0.34 & $0.334190  -3.81898\times10^{-5}  \imo$ & $0.334177-5.81446\times10^{-5}  \imo$ \\
%0.35 & $0.343422  -5.96741\times10^{-5}  \imo$ & $0.343554-8.85863\times10^{-5}  \imo$ \\
0.36 & $0.353692  -8.15278\times10^{-5}  \imo$ & $0.352876-1.32914\times10^{-4}  \imo$ \\
%0.37 & $0.362440  -1.31879\times10^{-4}  \imo$ & $0.362139-1.96121\times10^{-4}  \imo$ \\
0.38 & $0.371561  -1.99864\times10^{-4}  \imo$ & $0.371341-2.84203\times10^{-4}  \imo$ \\
%0.39 & $0.380643  -2.98943\times10^{-4}  \imo$ & $0.380480-4.03975\times10^{-4}  \imo$ \\
0.40 & $0.389695  -4.41105\times10^{-4}  \imo$ & $0.389556-5.62741\times10^{-4}  \imo$ \\
%0.41 & $0.398751  -6.39696\times10^{-4}  \imo$ & $0.398569-7.67867\times10^{-4}  \imo$ \\
0.42 & $0.407611  -8.88841\times10^{-4}  \imo$ & $0.407524-1.02633\times10^{-3}  \imo$ \\
\hline
\hline
\end{tabular}
}%
\caption{Schwarzschild (\(M=1\)) quasi-bound-state frequencies for a massive scalar field with \(\ell=1\). The WKB real parts were obtained using the Padé approximant (\(\widetilde{m}=\widetilde{n}=8\)) at the local minimum, while the imaginary parts were estimated using the Gamow approximation.  The results are compared with Frobenius values calculated using a continued fraction truncated at \(N=4000\).}
\label{tab:schwarzschildcompare}
\end{table}

\begin{figure*}
\resizebox{\linewidth}{!}{\includegraphics{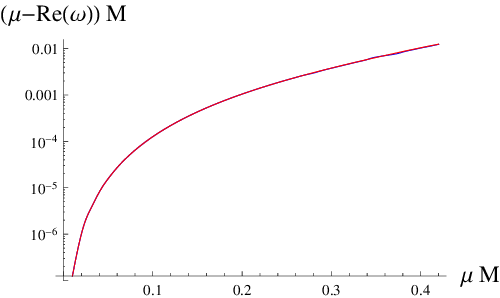}\includegraphics{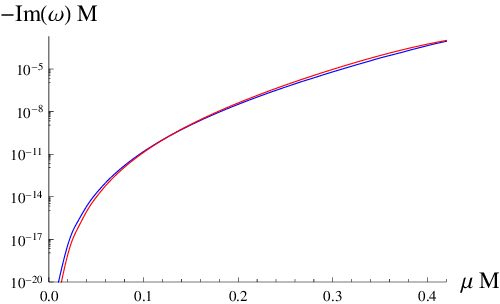}}
\caption{Binding energies \(\mu-\re{\omega}\) and decay rates \(-\im{\omega}\) for Schwarzschild (\(M=1\)) quasi-bound states of a massive scalar field with \(\ell=1\). The WKB approximations are shown in blue and the accurate values in red.}
\label{fig:complexmode}
\end{figure*}

For \(a=0\), \(\Lambda_{\ell m}=\ell(\ell+1)\), and the potential \eqref{eq:kerrpotential} becomes frequency independent and reduces to Eq.~\eqref{eq:potential} for the Schwarzschild black hole.
Therefore, the WKB quantization at the minimum and the Gamow calculation of the leakage through the barrier are directly applicable in the form described above.
Table~\ref{tab:schwarzschildcompare} compares the WKB results with accurate frequencies for the lowest Schwarzschild quasi-bound state. The agreement in the real part is very good over the displayed mass range. The imaginary part is more sensitive because it is exponentially small and is controlled by barrier penetration; nevertheless, the WKB calculation reproduces its scale and its variation with \(\mu\). The mass dependence is illustrated in Fig.~\ref{fig:complexmode}.

Unlike the Schwarzschild potential, the Kerr potential depends on the frequency through \(K(r)\) and \(\Lambda_{\ell m}\). We calculate \(\omega_R\) by solving Eq.~\eqref{eq:genericwkb} numerically, evaluating the Padé resummation of its right-hand side at the local minimum of the effective potential for each trial frequency.

However, when we use the WKB and Gamow formulas derived for frequency-independent potentials to calculate the imaginary part through Eq.~\eqref{eq:imagomegaexpanded}, we neglect additional terms involving derivatives of the potential with respect to the spectral parameter, such as \(\partial V/\partial\omega\). In particular, this approximation does not account for superradiant growth when \(0<\omega_R<ma/(2Mr_+)\). The regime in which the ordinary Gamow prescription fails cannot, however, be determined accurately from the WKB expansion~\eqref{eq:genericwkb} (see, e.g., the discussion in Ref.~\cite{Konoplya:2024vuj}). Nevertheless, the prescription gives a reasonable approximation on the decaying branch, reproducing the rapid suppression of the decay as the superradiant threshold is approached.

\begin{table}
\CQGonly{\flushright}
\PRDonly{\resizebox{\linewidth}{!}}{%
\begin{tabular}{|c|c|c|}
\hline
\hline
$a$ & WKB & Frobenius \\
\hline
%$-1.0^{\ast}$ & $0.295934-4.60464\times10^{-5}\imo$ & $0.296055-5.06380\times10^{-5}\imo$ \\
$-0.9$ & $0.296007-3.91913\times10^{-5}\imo$ & $0.296069-4.46936\times10^{-5}\imo$ \\
$-0.8$ & $0.296050-3.32877\times10^{-5}\imo$ & $0.296084-3.91953\times10^{-5}\imo$ \\
$-0.7$ & $0.296081-2.80923\times10^{-5}\imo$ & $0.296098-3.41286\times10^{-5}\imo$ \\
$-0.6$ & $0.296107-2.35115\times10^{-5}\imo$ & $0.296112-2.94776\times10^{-5}\imo$ \\
$-0.5$ & $0.296130-1.94892\times10^{-5}\imo$ & $0.296126-2.52262\times10^{-5}\imo$ \\
$-0.4$ & $0.296151-1.59802\times10^{-5}\imo$ & $0.296140-2.13588\times10^{-5}\imo$ \\
$-0.3$ & $0.296171-1.29437\times10^{-5}\imo$ & $0.296153-1.78612\times10^{-5}\imo$ \\
$-0.2$ & $0.296190-1.03414\times10^{-5}\imo$ & $0.296166-1.47198\times10^{-5}\imo$ \\
$-0.1$ & $0.296206-8.13439\times10^{-6}\imo$ & $0.296179-1.19220\times10^{-5}\imo$ \\
$0$    & $0.296219-6.27991\times10^{-6}\imo$ & $0.296192-9.45565\times10^{-6}\imo$ \\
$0.1$  & $0.296238-4.73116\times10^{-6}\imo$ & $0.296205-7.30865\times10^{-6}\imo$ \\
$0.2$  & $0.296204-3.50737\times10^{-6}\imo$ & $0.296217-5.46887\times10^{-6}\imo$ \\
$0.3$  & $0.296238-2.46473\times10^{-6}\imo$ & $0.296229-3.92380\times10^{-6}\imo$ \\
$0.4$  & $0.296249-1.65687\times10^{-6}\imo$ & $0.296241-2.66023\times10^{-6}\imo$ \\
$0.5$  & $0.296257-1.03863\times10^{-6}\imo$ & $0.296253-1.66395\times10^{-6}\imo$ \\
$0.6$  & $0.296267-5.84767\times10^{-7}\imo$ & $0.296264-9.19254\times10^{-7}\imo$ \\
$0.7$  & $0.296273-2.75013\times10^{-7}\imo$ & $0.296275-4.07992\times10^{-7}\imo$ \\
$0.8$  & $0.296285-8.97749\times10^{-8}\imo$ & $0.296285-1.07614\times10^{-7}\imo$ \\
$0.9$  & $0.296296-2.06549\times10^{-8}\imo$ & $0.296296+1.47776\times10^{-8}\imo$ \\
%$1.0^{\ast}$ & $0.296305-2.15923\times10^{-8}\imo$ & $0.296305+2.63745\times10^{-8}\imo$ \\
\hline
\hline
\end{tabular}
}
\caption{Kerr (\(M=1\)) frequencies of the fundamental quasi-bound state of a massive scalar field with \(\mu M=0.3\) and \(\ell=m=1\).  The case \(a/M=0.9\) lies in the superradiant regime \(\omega_R<m\Omega_H\), where \(\Omega_H=a/(2Mr_+)\). The WKB real parts were obtained by numerically solving the WKB equation with the Padé approximant (\(\widetilde{m}=\widetilde{n}=8\)) at the local minimum. The imaginary parts are calculated using the Gamow formula for a frequency-independent potential and do not include the superradiant horizon-flux correction. The results are compared with Frobenius values calculated using a continued fraction truncated at \(N=4000\).
}
\label{tab:kerrcompare}
\end{table}

Table~\ref{tab:kerrcompare} compares the WKB approximation with the accurate Frobenius values for \(\mu M=0.3\) and \(\ell=m=1\). At \(a=0\), the calculation reduces to the corresponding Schwarzschild result. The WKB and Frobenius real parts differ by less than about \(0.1\%\). The WKB imaginary part is less accurate and generally underestimates the magnitude of the decay rate.  For \(a/M=0.9\), the accurate Frobenius result has \(\im{\omega}>0\), as expected for a superradiant instability, whereas the WKB--Gamow imaginary part remains negative.

\begin{table*}
\resizebox{\linewidth}{!}{%
\begin{tabular}{|c|cc|cc|cc|cc|cc|}
\hline
\hline
& \multicolumn{2}{c|}{BS Fitting}
& \multicolumn{2}{c|}{BS Low}
& \multicolumn{2}{c|}{BS High}
& \multicolumn{2}{c|}{Numerical}
& \multicolumn{2}{c|}{WKB}\\
\cline{2-11}
$n$ & $\re{\omega_n}$ & $|\im{\omega_n}|$
& $\re{\omega_n}$ & $|\im{\omega_n}|$
& $\re{\omega_n}$ & $|\im{\omega_n}|$
& $\re{\omega_n}$ & $|\im{\omega_n}|$
& $\re{\omega_n}$ & $|\im{\omega_n}|$ \\
\hline
\multicolumn{11}{|c|}{$\ell=2$} \\
\hline
0 & $0.1060$ & $2.88\times10^{-9}$  & $0.1068$ & $1.01\times10^{-9}$  & $0.1091$ & $1.29\times10^{-9}$  & $0.1090$ & $1.24\times10^{-9}$ & $0.1090$ & $1.28\times10^{-9}$ \\
1 & $0.1530$ & $9.54\times10^{-8}$  & $0.1462$ & $3.16\times10^{-8}$  & $0.1485$ & $3.73\times10^{-8}$  & $0.1484$ & $3.95\times10^{-8}$ & $0.1485$ & $3.72\times10^{-8}$  \\
2 & $0.1970$ & $1.14\times10^{-6}$  & $0.1856$ & $4.13\times10^{-7}$  & $0.1879$ & $4.72\times10^{-7}$  & $0.1876$ & $5.47\times10^{-7}$ & $0.1879$ & $4.71\times10^{-7}$  \\
3 & $0.2377$ & $7.92\times10^{-6}$  & $0.2251$ & $3.47\times10^{-6}$  & $0.2274$ & $3.88\times10^{-6}$  & $0.2267$ & $4.85\times10^{-6}$ & $0.2272$ & $3.86\times10^{-6}$  \\
4 & $0.2756$ & $3.97\times10^{-5}$  & $0.2646$ & $2.28\times10^{-5}$  & $0.2668$ & $2.52\times10^{-5}$  & $0.2654$ & $3.23\times10^{-5}$ & $0.2668$ & $2.52\times10^{-5}$  \\
5 & $0.3114$ & $1.60\times10^{-4}$  & $0.3040$ & $1.40\times10^{-4}$  & $0.3063$ & $1.52\times10^{-4}$  & $0.3036$ & $1.72\times10^{-4}$ & $0.3062$ & $1.52\times10^{-4}$  \\
6 & $0.3453$ & $5.54\times10^{-4}$  & $0.3411$ & $5.51\times10^{-4}$  & $0.3441$ & $6.14\times10^{-4}$  & $0.3410$ & $7.30\times10^{-4}$ & $0.3457$ & $6.50\times10^{-4}$  \\
7 & $0.3778$ & $1.71\times10^{-3}$  & $0.3784$ & $2.00\times10^{-3}$  & $0.3752$ & $1.81\times10^{-3}$  & $0.3777$ & $2.30\times10^{-3}$ & $0.3851$ & $2.45\times10^{-3}$  \\
\hline
\multicolumn{11}{|c|}{$\ell=3$} \\
\hline
0  & $0.1474$ & $3.46\times10^{-13}$ & $0.1494$ & $1.34\times10^{-13}$ & $0.1506$ & $1.56\times10^{-13}$ & $0.1508$ & $1.52\times10^{-13}$ & $0.1508$ & $1.60\times10^{-13}$ \\
1  & $0.1949$ & $1.99\times10^{-11}$ & $0.1886$ & $6.70\times10^{-12}$ & $0.1901$ & $7.61\times10^{-12}$ & $0.1901$ & $7.76\times10^{-12}$ & $0.1901$ & $7.60\times10^{-12}$ \\
2  & $0.2422$ & $4.62\times10^{-10}$ & $0.2279$ & $1.37\times10^{-10}$ & $0.2294$ & $1.52\times10^{-10}$ & $0.2293$ & $1.62\times10^{-10}$ & $0.2293$ & $1.52\times10^{-10}$ \\
3  & $0.2869$ & $5.82\times10^{-9}$  & $0.2671$ & $1.71\times10^{-9}$  & $0.2686$ & $1.87\times10^{-9}$  & $0.2686$ & $2.08\times10^{-9}$  & $0.2686$ & $1.87\times10^{-9}$  \\
4  & $0.3292$ & $4.88\times10^{-8}$  & $0.3064$ & $1.54\times10^{-8}$  & $0.3079$ & $1.67\times10^{-8}$  & $0.3078$ & $1.93\times10^{-8}$  & $0.3079$ & $1.67\times10^{-8}$  \\
5  & $0.3694$ & $3.09\times10^{-7}$  & $0.3457$ & $1.10\times10^{-7}$  & $0.3472$ & $1.18\times10^{-7}$  & $0.3469$ & $1.42\times10^{-7}$  & $0.3472$ & $1.17\times10^{-7}$  \\
6  & $0.4078$ & $1.59\times10^{-6}$  & $0.3850$ & $6.54\times10^{-7}$  & $0.3865$ & $6.99\times10^{-7}$  & $0.3860$ & $8.79\times10^{-7}$  & $0.3865$ & $6.98\times10^{-7}$  \\
7  & $0.4447$ & $6.98\times10^{-6}$  & $0.4243$ & $3.45\times10^{-6}$  & $0.4257$ & $3.66\times10^{-6}$  & $0.4249$ & $4.72\times10^{-6}$  & $0.4258$ & $3.66\times10^{-6}$  \\
8  & $0.4802$ & $2.71\times10^{-5}$  & $0.4636$	& $1.83\times10^{-5}$  & $0.4650$ & $1.93\times10^{-5}$  & $0.4636$ & $2.25\times10^{-5}$  & $0.4650$ & $1.82\times10^{-5}$ \\
9  & $0.5146$ & $9.54\times10^{-5}$  & $0.5019$ & $7.32\times10^{-5}$  & $0.5075$ & $8.93\times10^{-5}$  & $0.5019$ & $9.64\times10^{-5}$  & $0.5043$ & $7.98\times10^{-5}$ \\
10 & $0.5479$ & $3.09\times10^{-4}$  & $0.5403$ & $2.83\times10^{-4}$  & $0.5404$ & $2.84\times10^{-4}$  & $0.5397$ & $3.63\times10^{-4}$  & $0.5436$ & $3.17\times10^{-4}$ \\
11 & $0.5803$ & $9.36\times10^{-4}$  & $0.5782$ & $1.03\times10^{-3}$  & $0.5768$ & $9.82\times10^{-4}$  & $0.5768$ & $1.15\times10^{-3}$  & $0.5829$ & $1.20\times10^{-3}$ \\

\hline
\hline
\end{tabular}
}%
\caption{Real and imaginary parts of the axial trapped-mode frequencies of uniform-density stars with $R/M=2.26$ for $\ell=2$ and $\ell=3$. The Bohr--Sommerfeld approximations and accurate numerical results of Table~I of Ref.~\cite{Volkel:2017ofl} are compared with the results of the present approach, in which the real parts are calculated using the Padé approximant (\(\widetilde{m}=\widetilde{n}=8\)) to the local WKB formula and the imaginary parts are estimated using the Gamow approximation.}
\label{tab:uniformdensitystar}
\end{table*}

Another example of the application of our approach is the calculation of the stellar trapped-mode frequencies presented in Table~\ref{tab:uniformdensitystar}. We compare our approximation with the results of Ref.~\cite{Volkel:2017ofl}, where the Bohr--Sommerfeld condition was employed instead of the local WKB formula. The ``BS Fitting'' frequencies were obtained by approximating the trapping well and the barrier with analytically tractable potentials. The ``BS Low'' results follow from a numerical evaluation of the leading Bohr--Sommerfeld condition and the Gamow formula, whereas ``BS High'' includes the leading higher-order correction to the Bohr--Sommerfeld condition for the real part. The ``Numerical'' column contains the frequencies obtained by solving the perturbation equation directly.

For the lower overtones, the WKB values of $\re{\omega_n}$ agree closely with the fully numerical results. The corresponding estimates of $|\im{\omega_n}|$ are comparable to those obtained with the higher-order Bohr--Sommerfeld rule because both methods use the same leading semiclassical description of tunnelling. Once the real frequency is known accurately, it fixes the turning points, the barrier action, and the well-normalization integral entering the Gamow decay rate. Conversely, even a small error in $\re{\omega_n}$ can produce a much larger relative error in the imaginary part because the transmission probability depends exponentially on the barrier action. At higher overtones, a fixed-order local WKB formula becomes less reliable. For smooth effective potentials, this loss of accuracy can be mitigated using the very-high-order WKB expansion and resummation procedure of Ref.~\cite{Konoplya:2026rjh}, which generates the WKB series to arbitrarily high order and resums it using Padé approximants. In the compact-star example, however, the discrepancy is not solely a finite-order WKB effect. The effective potential of a uniform-density star is discontinuous at the stellar surface, and higher trapped overtones extend farther from the local minimum and become increasingly sensitive to this discontinuity. Because the local WKB formula depends only on derivatives of the potential at the minimum, it cannot encode the surface discontinuity even when carried to higher orders, resulting in a systematic error in the real part of the frequency. The surface discontinuity is therefore the principal source of the discrepancy between the local WKB and higher-order Bohr--Sommerfeld results for the higher trapped overtones in Table~\ref{tab:uniformdensitystar}. Although the local formula is not the most appropriate approximation for this extreme example, it nevertheless yields quite accurate results for the lower overtones, whose real parts are governed primarily by the shape of the potential near its minimum.

Compared with the higher-order Bohr--Sommerfeld approach, the local WKB calculation of the real part avoids the repeated evaluation and differentiation of well-action integrals with energy-dependent turning points. It requires only the derivatives of the potential at its minimum and can consequently be implemented as a largely automatic algebraic procedure, including for non-rational effective potentials. We publicly share the Mathematica notebook implementing this procedure~\cite{WKBNotebook}. The leading barrier and well integrals are still required by the Gamow formula for the imaginary part. The accuracy of the local expansion must nevertheless be checked by increasing the WKB order and comparing neighboring resummations, particularly for modes that are not strongly localized near the minimum.

\section{Discussion}\label{sec:higher-order}

The Gamow expansion employed in the present manuscript gives the simplest semiclassical approximation to the imaginary part of the quasi-bound-state frequency, but generally provides only a rough estimate. One possible route to greater accuracy is to extend the calculation to higher WKB orders. In this section, we briefly discuss two possible extensions and the difficulties encountered when testing them.

\subsection{Higher-order approximation for tunnelling}

The Gamow result \eqref{eq:imagomegaexpanded} depends on two quantities: the tunnelling probability and the timescale between repeated encounters with the barrier. Both quantities can instead be estimated using the local higher-order WKB formula \eqref{eq:genericwkb}.

The analogue of \(I'(\omega_R^2)\) is obtained from the dependence of the WKB level at the minimum on the overtone number. The leading Bohr--Sommerfeld relation is
\begin{equation}
 I(\omega_R^2)=\pi\left(n+\frac12\right),
 \label{eq:actionquantization}
\end{equation}
where \(\omega_R\) is a function of the overtone number \(n\).
Differentiating this relation gives
\begin{equation}
 I'(\omega_R^2)\simeq\frac{\pi}{d\omega_R^2/dn}=\frac{\pi}{d\omega_R^2/dK}\Biggr|_{K=n+1/2}.
 \label{eq:effectiveIprime}
\end{equation}

Here \(\omega_R^2\) is calculated using the analytic formula \eqref{eq:genericwkb}, with its explicit dependence on \(K\). In this way, the higher-order WKB terms are included in the derivative, and the same Padé resummation can be used to calculate both \(\omega_R\) and \(I'(\omega_R^2)\).

Next, the WKB formula \eqref{eq:genericwkb} is evaluated at the maximum. After substituting the squared real frequency \(\omega_R^2\), we solve the formula for \(K\). The resulting value \(K_{\max}\) determines the
transmission probability:
\begin{equation}
 |T(\omega_R)|^2
 =
 \frac{1}{1+\exp(2\pi\imo K_{\max})}.
 \label{eq:peaktransmission}
\end{equation}
At leading order,
\begin{equation}
 \imo K_{\max}
 \simeq
 \frac{V_0-\omega_R^2}{\sqrt{-2V_2}},
 \label{eq:peakKleading}
\end{equation}
where \(V_0\) and \(V_2\) are evaluated at the maximum. At higher orders, all the derivatives of the potential entering Eq.~\eqref{eq:genericwkb} are evaluated at the maximum, and the resulting nonlinear equation is solved for the root continuously connected to \eqref{eq:peakKleading}. However, a straightforward Padé resummation does not provide a reliable prescription for this inversion (see Ref.~\cite{Konoplya:2019hlu} for details).

Accordingly, the transmission factor in the Gamow formula can be replaced by the higher-order WKB estimate
\begin{equation}
 e^{-2K(\omega_R^2)}\simeq|T(\omega_R)|^2 .
 \label{eq:effectiveK}
\end{equation}
This estimate uses the local shape of the peak rather than an integral over the entire forbidden region.

For a sufficiently opaque barrier, substituting
\eqref{eq:effectiveIprime} and \eqref{eq:effectiveK} into the Gamow formula
\eqref{eq:imagomegaexpanded} gives
\begin{equation}
 \omega\simeq
 \omega_R-\imo\frac{|T(\omega_R)|^2}{8\pi\omega_R}
 \frac{d\omega_R^2}{dn}.
 \label{eq:peakimagomegaexpanded}
\end{equation}

Thus, the calculation requires only two applications of the WKB formula. At the minimum, it gives \(\omega_R^2(n)\) and its derivative with respect to \(n\); at the maximum, it gives the transmission coefficient.

In our exploratory calculations for Schwarzschild quasi-bound states, this approach was less stable than the leading Gamow approximation and required careful selection of the WKB orders at the minimum and maximum. We therefore find the Gamow expansion more practical for estimating quasi-bound-state frequencies.

\subsection{Higher-order corrections to the action integrals}\label{sec:higher-order-Galtsov}

Another way to improve the Gamow approximation is to calculate higher-order corrections directly to the actions \(K(\omega_R^2)\) and \(I(\omega_R^2)\). Following Ref.~\cite{Galtsov:1991nwq}, we define
\begin{equation}
 Q^2(x)=\omega_R^2-V(x),
\end{equation}
and seek local solutions to Eq.~\eqref{eq:radial} in the form
\begin{equation}
 R(x)=q^{-1/2}(x)
 \exp\left(\mathord{\pm}\imo\intop^x q(z)\,dz\right).
 \label{eq:galtsovansatz}
\end{equation}
Substitution into the radial equation gives the exact condition
\begin{equation}
 q^2(x)
 =
 Q^2(x)+q^{1/2}(x)\frac{d^2}{dx^2}q^{-1/2}(x).
 \label{eq:galtsovequation}
\end{equation}
The usual WKB momentum \(Q(x)\) is the leading-order approximation to \(q(x)\).
Iterating Eq.~\eqref{eq:galtsovequation} gives
\begin{eqnarray}\label{eq:galtsovseries}
 q(x)&=&Q(x)+\frac{1}{2}Q^{-1/2}(x) \frac{d^2}{dx^2}Q^{-1/2}(x)
 \\\nonumber&&
 -\frac{1}{8}Q^{-1/2}(x)
 \frac{d^2}{dx^2}
 \left[
 Q^{-2}(x)\frac{d^2}{dx^2}Q^{-1/2}(x)
 \right]+\ldots\,.
\end{eqnarray}
Thus, higher-order corrections to an action are obtained by replacing its leading-order momentum with the corresponding truncation of Eq.~\eqref{eq:galtsovseries} and integrating term by term along an appropriate complex contour.

For example, in the forbidden region let
\begin{equation}
 q_0(r)=\sqrt{V(r)-\omega_R^2}.
\end{equation}
The barrier action in Eq.~\eqref{eq:radialactions} and its first correction can formally be written as
\begin{eqnarray}\label{eq:galtsovradialintegrand}
 K(\omega_R^2)&=&\intop_{r_1}^{r_2}\left(\frac{B(r)q_0(r)}{f(r)}
\PRDonly{%
 \right.
 \\\nonumber&&
 \left.
}
 +\frac{q_0^{-1/2}(r)}{2}\frac{d}{dr}\frac{f(r)}{B(r)}\frac{d q_0^{-1/2}(r)}{dr}\right)\,dr.
\end{eqnarray}
The first term gives the original barrier action. The second is the first correction, with the overall phase fixed by the analytic continuation from the oscillatory form \eqref{eq:galtsovansatz}. The well action is treated in the same way, using the branch that reduces to \(\sqrt{\omega_R^2-V(r)}\) between the two turning points. Once the corrected well action has been obtained, its derivative with respect to \(\omega_R^2\) gives the corresponding correction to \(I'(\omega_R^2)\).

The individual correction integrals cannot be evaluated directly on the real axis. Near a simple turning point \(r_t\),
\begin{equation}
 q_0(r)\propto (r-r_t)^{1/2}.
\end{equation}
Consequently, the second term in Eq.~\eqref{eq:galtsovradialintegrand} scales as \((r-r_t)^{-5/2}\). Its real-axis integral is therefore divergent. Every additional differentiation in \eqref{eq:galtsovseries} produces still stronger endpoint singularities. This divergence does not represent a physical divergence of the action. It shows that the derivative expansion is not uniform at a turning point and that its terms must not be integrated separately along a path ending there.

The integral must instead be evaluated along a contour in the complex \(r\) plane. The contour must satisfy four conditions: it must stay a finite distance from the turning points, enclose only the desired turning-point pair, avoid the poles of \(f^{-1}(r)\) and all other singularities, and maintain a single continuously tracked branch of every fractional power of \(q_0(r)\).

For the leading action, an upper semicircle connecting the two turning points is adequate because its endpoint singularities are integrable. It is not adequate for the correction in \eqref{eq:galtsovradialintegrand}, since the semicircle still terminates at the singular points. The corrected expression requires a closed contour with finite clearance around the entire cut. If the branch changes sign across the cut, the contour integral obeys the schematic relation
\begin{equation}
 \oint_{\mathcal C}\frac{B(r)q_0(r)}{f(r)}\,dr
 =
 \mathord{\pm}2\int_{r_1}^{r_2}
 \frac{B(r)q_0(r)}{f(r)}\,dr,
 \label{eq:contournormalization}
\end{equation}
up to the phase determined by whether the interval is allowed or forbidden. The sign, phase, and factor of two should be fixed by requiring the zeroth-order contour integral to reproduce the already known real action. Exactly the same contour, branch, and normalization must then be used for the corrected integrand.

There are two further complications. First, in the Gal'tsov construction the turning points are zeros of the modified quantity \(q^2\), rather than the zeros of the original \(Q^2\). Their positions therefore change with the
order of approximation. Keeping the original turning points while adding more singular correction terms is not a consistent higher-order procedure. Second, contour regularization makes a given truncated integral finite, but
does not by itself guarantee convergence with WKB order. The controlled quantity in the matrix WKB method is the deviation of the comparison equation from the exact equation. A useful numerical result requires this deviation to remain small on the whole contour.

In practice, one should first verify the leading contour integral, then add one correction at a time, track the corrected turning points and the analytic branch continuously, and vary the contour without crossing any singularity. A valid result should remain unchanged under such contour deformations and should stabilize as the order is increased.

We were unable to obtain stable results with this approach. If such a procedure exists, it appears to be more complicated than calculating the quasi-bound states directly by numerical shooting and matching solutions of the Riccati equation.

\section{Conclusions}\label{sec:conclusions}

In this work, we have developed a largely automatic semi-analytic method for calculating weakly damped modes supported by a smooth, nondegenerate minimum of an effective potential. The real part of the frequency is obtained from the higher-order local WKB expansion, supplemented by Padé resummation, whereas the imaginary part is estimated from the leading Gamow tunnelling probability.

An important advantage of our approach is its applicability to a broad range of spectral problems involving \emph{various} static and stationary compact objects, including:
\begin{itemize}
\item quasi-bound states of black holes,
\item quasi-bound states of wormholes,
\item trapped quasinormal modes of compact stars and regular gravastars,
\item bound states of compact stars and regular gravastars.
\end{itemize}
The approach is also efficient for configurations in which a trapping well is separated from both asymptotic regions by potential barriers.

We have compared the method with accurate continued-fraction results for massive scalar fields around Schwarzschild and Kerr black holes and with direct numerical results for axial trapped modes of a uniform-density star. The local WKB approximation reproduces the real frequencies with high accuracy whenever the modes are sufficiently localized near the potential minimum. The leading Gamow approximation is less accurate, but generally captures the scale and parameter dependence of the decay rate. For the Kerr black hole, the frequency dependence of the effective potential is treated approximately, and the method does not describe the transition from decay to superradiant growth. The WKB frequencies nevertheless provide useful initial guesses for accurate Frobenius calculations. We provide the Mathematica® notebook implementing the procedure~\cite{WKBNotebook}.

A principal advantage of the local WKB approach is that it requires only derivatives of the potential at its minimum and therefore avoids the higher-order evaluation of action integrals with energy-dependent turning points. It can consequently be applied to non-rational effective potentials with little case-specific modification. Extending the imaginary part systematically to higher WKB orders remains more difficult. Our attempts based on local peak transmission and contour-regularized corrections to the action integrals did not yield a stable procedure. Such a construction appears to be more complicated than directly solving the Riccati equation numerically. The present method nevertheless provides an efficient calculation of trapped-mode frequencies and a useful first estimate of their lifetimes.

\acknowledgments
The authors thank Sebastian H. Völkel and Oleksandr S. Stashko for useful discussions.
A. Z. was partially supported by Conselho Nacional de Desenvolvimento Científico e Tecnológico (CNPq).

\CQGonly{\bigskip}

\bibliography{bibliography}

\end{document}